\documentclass[aoas,MSNbibl,nameyear,seceqn,dvips]{arximspdf}
\usepackage{dcolumn}
\usepackage{graphicx}
\usepackage{textcomp}

\doi{10.1214/10-AOAS391}
\volume{5}
\issue{1}
\pubyear{2011}
\firstpage{400}
\lastpage{426}

\makeatletter
\newcolumntype{d}[1]{D{.}{.}{#1}}

\newcommand{\bX}{\mathbf{X}}
\newcommand{\bY}{\mathbf{Y}}
\newcommand{\nn}{\nonumber}
\newcommand{\trans}{^{\mathrm{T}}}
\newcommand{\blambda}{\bolds\lambda}
\newcommand{\btheta}{\bolds\theta}
\newcommand{\bdelta}{\bolds\delta}
\newcommand{\balpha}{\bolds\alpha}
\newcommand{\bphi}{\bolds\phi}
\newcommand{\bpsi}{\bolds\psi}
\newcommand{\bbeta}{\bolds\beta}
\newcommand{\bH}{\mathbf{H}}
\newcommand{\bA}{\mathbf{A}}
\newcommand{\bB}{\mathbf{B}}
\newcommand{\bI}{\mathbf{I}}
\newcommand{\bL}{\mathbf{L}}
\newcommand{\bR}{\mathbf{R}}
\newcommand{\bT}{\mathbf{T}}
\newcommand{\bW}{\mathbf{W}}

\makeatother

\begin{document}
\begin{frontmatter}

\title{HIV dynamics and natural history studies:
Joint modeling with doubly interval-censored event time~and infrequent
longitudinal data\thanksref{TT1}}

\runtitle{HIV dynamics and natural history studies}

\begin{aug}
\author[A]{\fnms{Li} \snm{Su}\corref{}\ead[label=e1]{li.su@mrc-bsu.cam.ac.uk}}
\and
\author[B]{\fnms{Joseph W.} \snm{Hogan}\ead[label=e2]{jhogan@stat.brown.edu}}

\runauthor{L. Su and J. W. Hogan}

\affiliation{MRC Biostatistics Unit and Brown University}

\address[A]{MRC Biostatistics Unit\\
Robinson Way\\
Cambridge CB$2$ $0$SR\\
UK\\
\printead{e1}} %adresu isvedimo komanda gale!

\address[B]{Center for Statistical Sciences\\
Department of Community Health\\
Brown University\\
Box G-S121-7\\
Providence, Rhode Island 02912\\
USA\\
\printead{e2}}

\end{aug}

\thankstext{TT1}{Supported by NIH Grants R01-AI-50505, R01-HL-79457
and Grant U.1052.00.009 from the Medical Research Council (UK).}

% HISTORY:
\received{\smonth{10} \syear{2009}}
\revised{\smonth{7} \syear{2010}}

% ABSTRACT
%
\begin{abstract}
Hepatitis C virus (HCV) coinfection has become one of the most
challenging clinical situations to manage in HIV-infected patients.
Recently the effect of HCV coinfection on HIV dynamics following
initiation of highly active antiretroviral therapy (HAART) has drawn
considerable attention. Post-HAART HIV dynamics are commonly studied in
short-term clinical trials with frequent data collection design. For
example, the elimination process of plasma virus during treatment is
closely monitored with daily assessments in viral dynamics studies of
AIDS clinical trials. In this article instead we use infrequent cohort
data from long-term natural history studies and develop a model for
characterizing post-HAART HIV dynamics and their associations with HCV
coinfection. Specifically, we propose a joint model for doubly
interval-censored data for the time between HAART initiation and viral
suppression, and the longitudinal CD4 count measurements relative to
the viral suppression.
Inference is accomplished using a fully Bayesian approach. Doubly
interval-censored data are modeled semiparametrically by Dirichlet
process priors and Bayesian penalized splines are used for modeling
population-level and individual-level mean CD4 count profiles. We use
the proposed methods and data from the HIV Epidemiology Research Study
(HERS) to investigate the effect of HCV coinfection on the response to HAART.
\end{abstract}

% KEYWORDS
%

\end{frontmatter}

%s1 ###
\section{Introduction}
\label{introduction2}

%s1.1 ###
\subsection{HIV dynamics following initiation of antiviral
therapy}\label{hivdynamicsintro}

The wide-spread use of highly active antiretroviral therapies
(HAART) against HIV~in~the United States has resulted in
reducing the burden of HIV-related morbidity and mortality [\citet
{jacobson2004}]. HIV dynamics following HAART are usually studied in
short-term clinical trials with frequent data collection design. For
example, in viral dynamics studies of AIDS clinical trials the
elimination process of plasma virus after treatment is closely
monitored with daily measurements, which has led to a new understanding
of the pathogenesis of HIV infection and provides guidance for treating
AIDS patients and evaluating antiviral therapies [\citet
{wu2005}]. Here
in this article HIV dynamics refer to a two-part response to HAART:
viral suppression and concurrent or subsequent immune reconstitution.
In clinical practice, the virus is considered suppressed when plasma
HIV RNA (viral load) is below a lower limit of detection; the degree of
immune reconstitution is commonly measured by the change of CD4$+$
lymphocyte cell count (CD4 count) after HAART initiation.

It is well known that CD4$+$ lymphocyte cells are targets of HIV and
their abundance declines after HIV infection. Investigators have
studied the association between viral load and CD4 count during HAART
treatment and, in general, they are negatively correlated [\citet
{lederman1998}; \citet{liang2003}]. Longitudinal data on these
markers have
been analyzed separately, particularly by using random-effects models.
Recently, bivariate linear mixed models were proposed to jointly model
viral load and CD4 count by incorporating correlated random effects.
These models were specified in terms of concurrent association between
viral load and CD4 count [\citet{thi2005}; \citet
{pantazis2005}]. However, a~natural time ordering for virologic and
immunologic response to HAART (or any antiviral therapy) is often
observed: when a patient begins a successful HAART regimen, viral
replication is usually inhibited first, leading to a decrease in viral
load; then, CD4 count often increases as the immune system begins to
recover. Consequently, increase in CD4
count is thought to depend on the degree of viral suppression; it may
be slower to respond than viral load or it may not increase at all if
the virus is not suppressed [\citet{jacobson2004}]. Therefore, it would
be advantageous to acknowledge these common sequential changes of viral
load and CD4 count when modeling post-HAART HIV dynamics.

%s1.2 ###
\subsection{Coinfection with Hepatitis C virus and HIV dynamics}\label
{hcvintro}

Hepatitis C virus (HCV) coinfection is estimated to occur
in $30\%$ of HIV-infected patients in the United States and has become
one of the most challenging clinical situations to manage in
HIV-infected patients [\citet{sherman2002}]. Several studies
have suggested that HCV serostatus is not associated with the virologic
response to HAART [\citet{greub2000}; \citet{rock2005}].
However, the evidence
for immunologic response is conflicting.
Some studies have shown that {HIV--HCV} coinfected patients have
a blunted immunologic response to HAART, compared to
those with HIV infection alone, although others
have found comparable degrees of immune reconstitution
in persons with HIV--HCV coinfection [\citet
{miller2005}; \citet{Stebbing2005}; \citet{Rockstroh2006};
\citet{Sullivan2006}]. A primary
motivation of our model is to investigate the effect of HCV coinfection
on post-HAART HIV dynamics using cohort data from natural history
studies. We focus on two important questions: ($1$) Do HCV-negative
patients have shorter time from HAART initiation to viral suppression?
($2$) Do HCV-negative patients have better immune reconstitution at the
time of viral suppression?
Note that in the second question the sequential nature of the virologic
and immunologic response to HAART is emphasized.

%s1.3 ###
\subsection{HIV natural history studies and the HERS}\label{hersintro}

Because the incidence of clinical progression to AIDS fell rapidly following
the widespread introduction of HAART in 1997, long-term clinical trials
in patients
with HIV become time-consuming and expensive [\citet{mocroft2006}].
Currently, natural history studies are the major source of knowledge
about the HIV epidemic and the full treatment effect of HAART over the
long term. For example, studies such as Multicenter AIDS Cohort Study
(MACS), Women's Interagency HIV Study (WIHS)
and Swiss HIV Cohort Study (SHCS) have played important roles in
understanding the science of HIV, the AIDS epidemic and the effects of
therapy [\citet{kaslow1987}; \citet{leder1994}; \citet
{barkan1998}]. In HIV natural
history studies, HIV viral load and CD4 count are usually measured with
wide intervals (e.g., every $6$ months approximately). Therefore, for
some event time of scientific interest, for example, the time between
HAART initiation and viral suppression, both the time origin (HAART
initiation) and the failure event (viral suppression) could be
interval-censored. This situation is referred to as `doubly
interval-censored data' in the literature. In fact, the statistical
research on doubly interval-censored data was primarily motivated by
scientific questions in HIV research, for example, modeling `AIDS
incubation time' between HIV infection and the onset of AIDS
[\citet{Gruttola1989}; \citet{Sun2006}]. Both
nonparametric and semiparametric methods have been proposed for the
estimation of the distribution function of the AIDS incubation time and
its regression analysis. A comprehensive review on the analysis of
doubly interval-censored data can be found in \citet{Sun2006}.

The HIV Epidemiology Research Study (HERS) is a multi-site longitudinal
cohort study of HIV natural history in women between $1993$ and $2001$
[\citet{Smith1997}]. At baseline between $1993$ and $1995$ the study
enrolled $871$ HIV-seropositive women and $439$ HIV-seronegative women
at high risk for HIV infection. Participants were scheduled for
approximately a $6$-year follow-up, where a variety of clinical,
behavioral and sociologic outcomes were recorded approximately every
$6$ months and measurements correspond to dates. The top part of
Table~\ref{HERSbaseline} gives selected baseline characteristics of
the~1310 study participants; more details can be found in \citet
{Smith1997}. Quantification of HIV RNA viral load was performed using a
branched-DNA (B-DNA) signal amplification assay with the detection
limit at $50$ copies/ml and flow cytometry from whole blood was used to
determine CD4 counts at each visit. All participants were HAART-naive
at baseline. During the study $382$ participants reported HAART use
based on information gathered during in-person interviews. Because
assessments were scheduled to be carried out every $6$ months and
participants were only asked about whether they were on HAART during
the last 6 months, exact dates for HAART initiation are not available.
The analysis in Section~\ref{analysis2} includes 374 women with HAART
use who had HIV sero-conversion before baseline and baseline HCV
coinfection information. Some characteristics of these 374 women are
presented at the bottom of Table~\ref{HERSbaseline}.

%
%t1 ###
\begin{table}
\tabcolsep=0pt
\caption{Selected characteristics of the 1310 HERS women (top) and the
374 HERS women included in the analysis (bottom) in Section~\protect\ref{analysis2}}
\label{HERSbaseline}
\begin{tabular*}{\textwidth}{@{\extracolsep{\fill}}ld{2.6}d{2.6}c@{}}
\hline
&\multicolumn{1}{c}{\textbf{HIV-positive}}& \multicolumn
{1}{c@{}}{\textbf{HIV-negative}}\\
&\multicolumn{1}{c}{$\bolds{(N=871)}$} & \multicolumn
{1}{c@{}}{$\bolds{(N=439)}$}\\
\hline
Median age at enrollment& 35.0 & 34.5 \\
Age range at enrollment & 16.4\mbox{--}55.2 & 16.6\mbox{--}56.0 \\
Injection drug user at enrollment ($\%$) & 25.1 & 26.4\\
CD4 count at enrollment ($\%$) & & \\
 \quad $<$200 &17.1& 0.0 \\
 \quad 200--499 &50.7 & 1.7\\
 \quad $\ge$500 &32.2 &98.3\\
HCV antibody test at enrollment ($\%$)& & \\
 \quad Positive & 60.3& 47.8\\
 \quad Negative & 38.8& 50.8\\
 \quad Missing & 0.9 & 1.4\\
[6pt]
&\multicolumn{1}{c}{HCV-positive}& \multicolumn{1}{c@{}}{HCV-negative}\\
&\multicolumn{1}{c}{($N=208$)} & \multicolumn{1}{c@{}}{($N=166$)}\\
[3pt]
Median follow-up time (months) & 67.3 & 71.0 \\
Median age at enrollment & 36.7& 33.1 \\
Age range at enrollment & 21.2\mbox{--}55.0 & 19.0\mbox{--}55.2 \\
Injection drug user at enrollment ($\%$) & 29.8 & 2.4\\
Ever on antiviral treatment before 1996 ($\%$)& 57.2 & 62.1 \\
CD4 count before first reported HAART use ($\%$) & & \\
 \quad $<$200 &34.6& 36.8 \\
 \quad 200--499 &52.9 & 45.8\\
 \quad $\ge$500 &12.5 &17.5\\
\hline
\end{tabular*}
\vspace{-10pt}
\end{table}

%
%f1 ###
\begin{figure}

\includegraphics{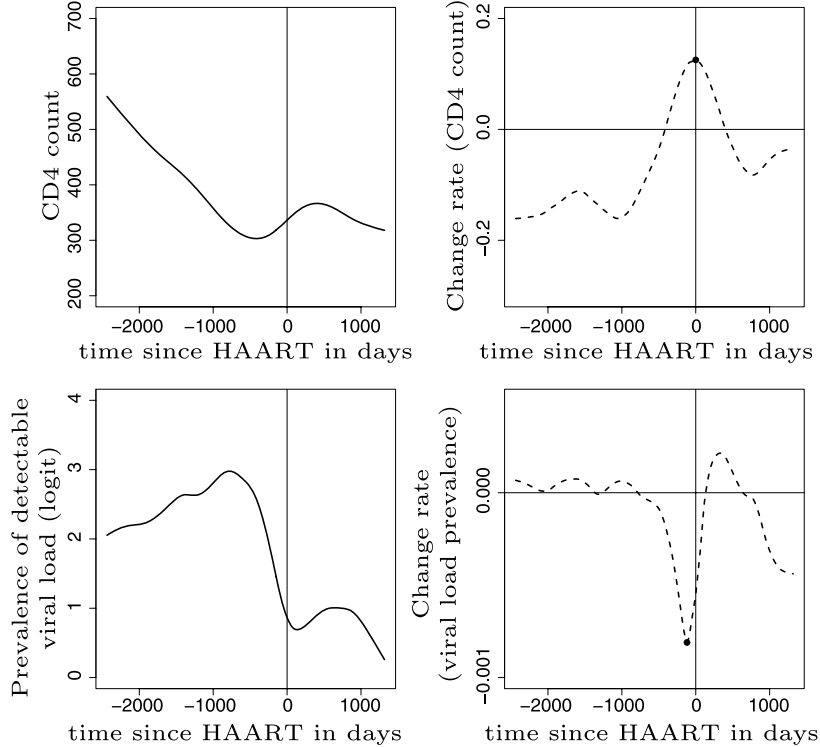}
\vspace{-5pt}
\caption{Top panels: smoothing spline fit and the corresponding
derivative (change rate) curve for average CD4 count since reported
HAART initiation in the HERS cohort; bottom panels: smoothing spline
fit and the corresponding derivative (change rate) curve for the
prevalence of detectable viral load ($\ge$50 copies/ml) since reported
HAART initiation in the HERS cohort; solid lines: smoothing spline
fits; dashed line: derivative curves of the smoothing spline fits;
black dots: maximum of the increasing rate for average CD4 count and
maximum of the decreasing rate of viral load prevalence.}\label{empinew}
\vspace{-10pt}
\end{figure}

Figure~\ref{empinew} shows smoothing spline fits and the corresponding
derivative (change rate) curves for average CD4 count and the
prevalence of detectable viral load for the 374 HERS women, where the
measurement times are cente\-red such that time $0$ represents the
earliest visit with HAART information reported. The left panels
indicate that the increasing trend for average CD4 count started later
than the decreasing trend for viral load prevalence, but this
phenomenon is probably not related to HAART because the starting times
for these trends are 1--2 years before the reported HAART initiation
time. It might be more useful to examine the change rates for average
CD4 count and viral load prevalence to assess the effectiveness of
HAART. In fact, the right panels of Figure~\ref{empinew} show that the
maximum decreasing rate for viral load prevalence occurred earlier
(around 4~months before repor\-ted HAART initiation) than the maximum
increasing rate for average CD4 count (around the reported HAART
initiation), which suggests the possible sequential relationship in
post-HAART HIV dynamics discussed in Section~\ref{hivdynamicsintro}.

%
%f2 ###
\begin{figure}
\vspace{-5pt}
\includegraphics{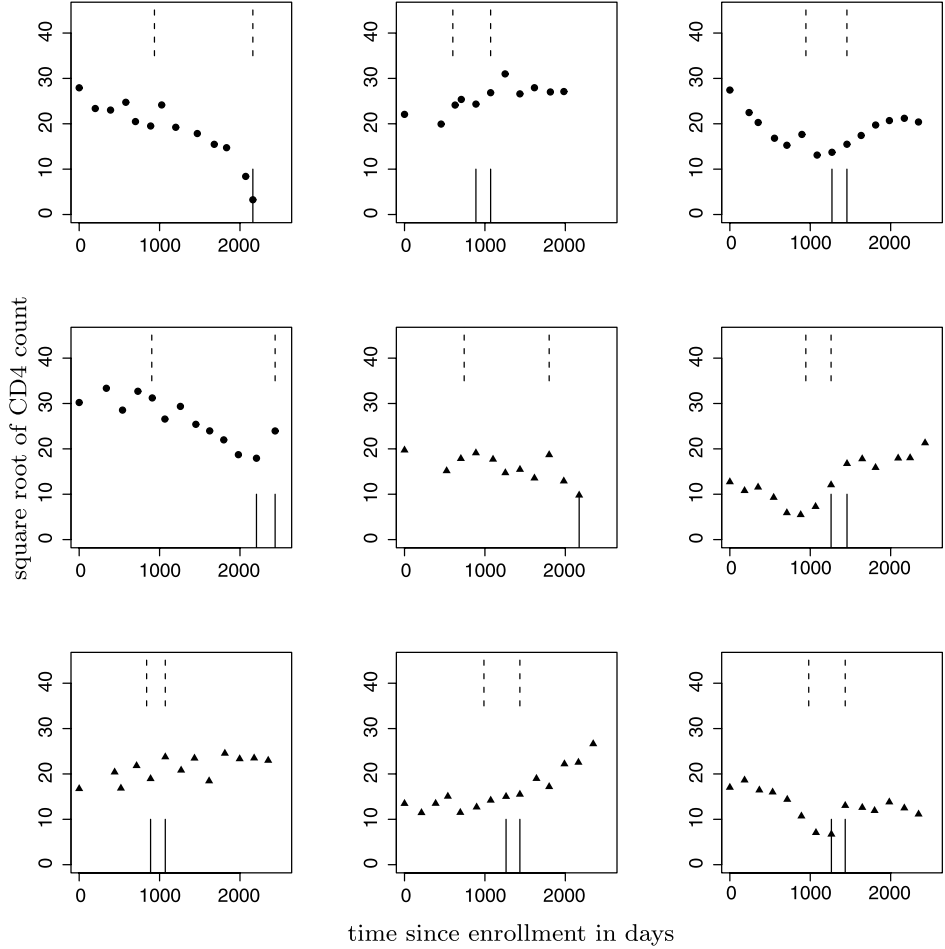}
\vspace{-20pt}
\caption{CD4 counts (on square root scale) and censoring intervals for
9 selected HERS women; dotted line: censoring intervals for HAART
initiation; solid line: censoring intervals for viral suppression
following HAART; circles represent the data from HCV-positive group and
triangles represent the data from HCV-negative group. }\label{selectedsub}
\vspace{-10pt}
\end{figure}

%s1.4 ###
\subsection{Modeling post-HAART HIV dynamics in the HERS}

Our objective is to develop a model for the joint distribution of the
time from HAART initiation to viral suppression, and the longitudinal
CD4 counts relative to the viral suppression time following HAART. As
discussed in Section~\ref{hersintro}, the time from HAART initiation
to viral suppression is doubly interval-censored. Specifically,
considering the reporting bias for HAART initiation, we define the
right endpoint of its corresponding censoring interval to be the first
visit of reported HAART use and the definition for the left endpoint is
based on assumptions about the earliest possible time of HAART
initiation in the HERS cohort. Further, viral suppression following
HAART can be either interval-censored or right-censored. Details can be
found in Section~\ref{analysis2}.

%
%f3 ###
\begin{figure}[b]

\includegraphics{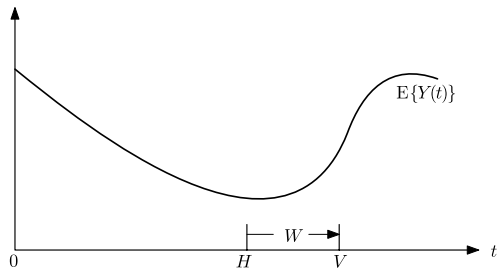}

\caption{A scheme of the variables of interest under an idealized situation
for post-HAART HIV dynamics: $0$ represents enrollment, $t$ indexes the
time since enrollment, $H$ is HAART initiation time, $V$~is viral
suppression time following HAART, $W$ is the time from HAART initiation
to viral suppression, and $Y(t_1),Y(t_2),\ldots, Y(t_n)$ are CD4 count
measurements with their expectations represented by the curve. }\label{scheme}
\vspace{-10pt}
\end{figure}

Figure~\ref{selectedsub} shows CD4 counts and corresponding censoring
intervals of HAART initiation and viral suppression following HAART for
selected HERS women. As seen in the top left panel of Figure~\ref
{selectedsub}, viral suppression after HAART can be right-censored due
to participant dropout, death and/or study end. Similarly, participants
could have incomplete CD4 count measurements for $12$ scheduled
follow-up visits. However, because we focus on the subpopulation of
HAART users in the HERS cohort, the missingness rate is relatively low
compared to the whole HERS population; $90.64\%$ of the $374$ women in
our analysis had at least $10$ visits. Therefore, for the HERS analysis
in Section~\ref{analysis2}, we assume that the missing data mechanism
is missingness at random [\citet{little2002}]. Given that the
parameters for modeling the missing data mechanism and the outcomes are
distinct and they have independent priors, the missing data are then
ignorable when making posterior inference about the outcomes.

The remainder of the article is organized as follows. In Section~\ref
{model2} we specify the joint model for doubly interval-censored event
time and longitudinal CD4 count data. Section~\ref{estimation2}
describes the posterior inference and gives full conditional
distributions for Gibbs steps. We use
the model to analyze the HERS data for investigating the HCV
coinfection problem introduced in Section~\ref{hcvintro}, and present
the results in
Section~\ref{analysis2}. The conclusion and some discussion are given
in Section~\ref{conclusion2}.

%s2 ###
\section{A model for post-HAART HIV dynamics}
\label{model2}

%s2.1 ###
\subsection{Model under an idealized situation}

Our goal is to model the joint distribution of the time from HAART
initiation to viral suppression and the longitudinal CD4 counts.
Figure~\ref{scheme} is a schematic illustration of the variables of
interest under an idealized situation.
Let $t$ $(t\ge0)$ denote the time since enrollment and let $H$ and $V$
represent the time from enrollment to HAART initiation and the time
from enrollment to viral suppression after HAART, respectively. By
definition, $V>H$ and $W=V-H$ is the time from HAART initiation to
viral suppression. Further, $Y(t_1),Y(t_2),\ldots, Y(t_n)$ are CD4
count measurements taken at time points $t_1<\cdots<t_n$. Throughout
this article, the time points $t_1<\cdots<t_n$ are assumed to be
noninformative and fixed by study design. Let $\bX$ denote covariates,
for example, the baseline HCV serostatus. The joint density of $W$ and
$Y(t_1),Y(t_2),\ldots, Y(t_n)$ given $\bX$, $H$ and $t_1, \ldots,
t_n$ can be written as
%
%
%e2.1 ###
\begin{eqnarray}\label{simple2}
&& \quad  p(w, y_1, y_2,\ldots, y_n \vert\bX, h, t_1, \ldots, t_n)
\nonumber
\\[-8pt]
\\[-8pt]
%&=& p_1(w\vert\bX, h, t_1, \ldots, t_n) p_2( y_1, y_2, \ldots,y_n\vert
&& \quad  \qquad = p(w\vert\bX, h)p\{ y_1, y_2, \ldots,y_n\vert\bX, t_1-(h+w),
\ldots,
t_n-(h+w)\}. \nn
\end{eqnarray}
The conditioning on $H$ is because we are not interested in the
marginal distribution of $H$ and the observed $H=h$ is only used as the
time origin for~$W$.

The factorization in (\ref{simple2}) is based on the sequential
relationship in post-HAART dynamics. When HAART regimen is successful
in suppressing the virus, we are able to obtain $W$, the time from
HAART initiation to viral suppression. As mentioned in Section~\ref
{hivdynamicsintro}, there is a time ordering of virologic response and
immunologic response to HAART.
Because of this sequential relationship of virologic and immunologic
response as well as the large between-individual heterogeneity in terms
of the ability to suppress viral replication, the time to suppression
and the durability of suppression,
we believe that the mean CD4 count profiles from different individuals
are more comparable after realigning measurement times by their
individual viral suppression times following HAART. Therefore, we
assume that the distribution of $Y(t_1),Y(t_2),\ldots, Y(t_n)$ given
$\bX$ depends on $H$ and $W$ only through a change in the time origin
for the measurement times $t_1, \ldots, t_n$.
This is similar to \textit{curve registration}, a method originated in
the functional data analysis literature [\citet{ramsay1998}] for dealing
with the situations where the rigid metric of physical time for real
life systems is not directly relevant to internal dynamics. For
example, the timing variation of salient features of individual puberty
growth curves (e.g., time of puberty growth onset, time of peak
velocity of puberty growth) can result in the distortion of population
growth curves [\citet{ramsay2005}]. Likewise, in our case, simply
averaging individual CD4 count profiles along the time since enrollment
($t$) or the time since HAART initiation ($H$) can attenuate the true
population immunologic response profile following HAART. Because viral
suppression is the main driving force of immune reconstitution
[\citet
{jacobson2004}], it is sensible to center the time scale at individual
viral suppression times ($V=H+W$) in order to describe the trends in
immune reconstitution at the population level.

However, as mentioned in Section~\ref{hersintro}, $W$ can be doubly
interval-censored in HIV natural history studies, which presents a
challenge in making inferences about the density in (\ref{simple2}).
In fact, for $p\{ y_1, y_2, \ldots,y_n\vert\bX, t_1-(h+w), \ldots,
t_n-(h+w)\}$, we are faced with a situation similar to the missing or
interval-censored covariate problem in generalized linear model
literature [\citet{chen2005}; \citet{calle2005}]. To
accommodate this situation, we will extend the semiparametric Bayesian
approach in \citet{calle2005} by modeling $H$ and $W$
simultaneously. Note that here we model the observed $H$ only for
taking into account the uncertainty in the time origin of $W$; we do
not intend to make inference about the marginal distribution for HAART
initiation time, which requests the right-censored data from those
participants who did not initiate HAART during the study. This is
different from the AIDS incubation time problem which motivated the
research in doubly interval-censored data, where both HIV infection
time and AIDS incubation time are of interest and HIV infection time
can be right-censored [\citet{Gruttola1989}]. Moreover, for the
HERS cohort, HAART was not available before 1996; therefore, when HAART
initiation time is of scientific interest, it is not valid to use
enrollment as the time origin because all HERS women were not at risk
for HAART initiation between enrollment and 1996. However, for the
purpose of accommodating uncertainty for the time origin of $W$, we can
still use the observed censoring intervals for $H$ with enrollment as
their time origin.

In the following sections, we present the details of the proposed joint
model for the HERS data.

%s2.2 ###
\subsection{Model with doubly interval-censored data}
%s2.2.1 ###
\subsubsection{Observed data}
Recall that all HERS women were HAART-naive at baseline. For those who
initiated HAART during follow-up,
let $H$ be a positive random variable representing the time from
enrollment to HAART initiation. Participants were monitored only
periodically, and at each follow-up visit they only reported whether
they were on HAART treatment since the last visit. Hence, the true
value for $H$ is only known to lie within an interval $(L^H, R^H]$,
where $L^H$ is the time of the visit preceding HAART initiation and
$R^H$ is the time of the first visit at which HAART use is reported.

Let $V$ be the time from enrollment to viral suppression following
HAART initiation. By definition, $V>H$. For those whose viral load has
been suppressed, $V$ is observed to be in an interval $(L^V, R^V]$,
where $L^V$ and $R^V$ are defined similarly as $L^H$ and $R^H$. For
those whose viral load was not suppressed during follow-up, $V \in
(L^V, +\infty)$, which corresponds to right censoring of $V$.
Because right censoring can be treated as a special case of interval
censoring with $R^V=+\infty$, we simply write $V \in(L^V, R^V]$. The
time between HAART initiation and viral suppression is $W=V-H$. At a
given value for $H$, $(L^H, R^H]$ and $(L^V, R^V]$ can overlap because
virus suppression can occur quickly after HAART but before the next
visit; therefore, $W \in( \operatorname{max}(0,L^V-H), R^V-H]$.

Further, we observe CD4 counts $\bY= (Y_{1},\ldots,Y_{n})\trans$ at
time points $t_{1},\ldots, t_{n}$, which can be different across
individuals and $\bX$ is the covariate that includes baseline HCV
coinfection status $Z$, where $Z \in\{0,1\}$ indicates positivity of
HCV antibody.

In summary, the observed data for a HAART user in the HERS cohort
consist of the observed CD4 counts $\bY$, the covariate $\bX$, the
observation times $t_{1},\ldots, t_{n}$ and the intervals $(L^H, R^H]$,
$(L^V, R^V]$ that respectively include HAART initiation time $H$ and
viral suppression time $V$.

%s2.2.2 ###
\subsubsection{Noninformative assumption for interval-censoring}
The joint density for the above observed data and the unobserved $H$
and $W$ can be written as
%
%
%e2.2 ###
\begin{eqnarray}\label{jointpdf}
&& p(l^H, r^H,l^V, r^V, h, w, \mathbf{y}  \vert\bX, t_{1},\ldots,
t_{n})\nonumber\\
&& \qquad =p_0(l^H, r^H, l^V, r^V \vert\bX) p_1( h \vert\bX,l^H, r^H, l^V,
r^V)\nonumber
\\[-8pt]
\\[-8pt]
&& \qquad  \quad {}\times p_2( w \vert\bX, h, l^H, r^H, l^V, r^V) \nonumber\\
&& \qquad  \quad {}\times p_3\{
\mathbf{y}
\vert\bX, t_{1}-(h+w),\ldots, t_{n}-(h+w), l^H, r^H, l^V, r^V\}.
\nonumber
\end{eqnarray}
%
%Our interest in modeling the effect of HCV coinfection on HIV dynamics
%is in $\blambda^W$ and $\btheta$, whereas $\bdelta$ and $\blambda^H$
%are specified only for the purpose of characterizing the uncertainty
%of the time origin of $W$.

Denote the cumulative distribution function (CDF) of $H$ given $\bX$ by
$G^H(h \vert\bX; \blambda^H)$, and the CDF of $W$ given $\bX$ by $G^W(w
\vert\bX; \blambda^W)$.
The corresponding probability density functions (PDF) are $g^H(h \vert
\bX; \blambda^H)$ and $g^W(w \vert\bX;\break \blambda^W)$, respectively. We
assume that the censoring of $H$ and $W$ occurs noninformatively
[\citet
{oller2004}; \citet{calle2005}], in the following sense:

\begin{enumerate}[(b)]
\item[(a)] $(L^H, R^H,L^V, R^V)$ provide no additional information
about $\bY$ when $H$ and $W$ are exactly observed. That is, the
conditional density of $\bY$ given $(\bX, H, W, t_{1},\ldots, t_{n})$
and $(L^H, R^H, L^V, R^V)$ does not depend on $(L^H, R^H, L^V, R^V)$:
\begin{eqnarray}
&& p_3\{\mathbf{y}\vert\bX, t_{1}-(h+w),\ldots, t_{n}-(h+w), l^H,
r^H, l^V,
r^V\} \nn\\
&& \qquad =p_3\{\mathbf{y}\vert\bX, t_{1}-(h+w),\ldots, t_{n}-(h+w); \btheta
\}.\nn
\end{eqnarray}
\item[(b)] The only information about $H$ and $W$ provided by the
observed censoring intervals is that $ (L^H, R^H]$, $(L^V, R^V]$
contain $H$ and $V=H+W$, respectively. That is, the conditional density
of $H$ given $\bX$ and $ (L^H, R^H]$ satisfies
%
%
%e2.3 ###
\begin{eqnarray}\label{truncatedH}
&& p_1(h \vert\bX, l^H, r^H, l^V, r^V)\nonumber
\\[-8pt]
\\[-8pt]
&& \qquad =\frac{g^H( h\vert\bX; \blambda^H)}{G^H(r^H\vert\bX; \blambda^H)-
G^H(l^H\vert\bX; \blambda^H)},\nn
\end{eqnarray}
which corresponds to the density of $H$ given $\bX$ truncated in $
(L^H, R^H]$. Similarly, the conditional density of $W$ given $\bX$, $H$
and $ (L^V, R^V]$ is
%
%
%e2.4 ###
\begin{eqnarray}\label{truncatedW}
&& p_2(w \vert\bX,h, l^H, r^H,l^V, r^V) \nonumber
\\[-8pt]
\\[-8pt]
&& \qquad =\frac{g^W( w\vert\bX; \blambda^W)}{G^W(r^V-h\vert\bX; \blambda^W)
-G^W(\operatorname{max}(0, l^V-h)\vert\bX; \blambda^W)},\nonumber
\end{eqnarray}
the truncated density $g^W(w \vert\bX;\blambda^W)$ in the interval
$(\operatorname{max}(0,L^V-H), R^V-H]$.
We denote (\ref{truncatedH}) by $g_T^H(h\vert\bX, l^H, r^H;\blambda^H)$
and (\ref{truncatedW}) by $g_T^W(w\vert\bX, h, l^V, r^V;\break \blambda^W)$,
where the subscript $T$ stands for `truncated' density.
%scheduled study observation times, we assume that the censoring is not
%differential by baseline HCV serostatus, i.e., $(L^H, R^H, L^V, R^V)$
%are independent of $Z$,
%p_0(l^H, r^H,l^V, r^V \vert z, t_{1},\ldots, t_{n}; \bdelta)= p_0( l^H,
%r^H,l^V, r^V \vert t_{1},\ldots, t_{n};\bdelta).
\end{enumerate}

Given these noninformative conditions, the joint density in (\ref
{jointpdf}) can be simplified as
%
%
%e2.5 ###
\begin{eqnarray}
&&p(l^H, r^H,l^V, r^V, h, w, \mathbf{y}, \vert\bX, t_{1},\ldots,
t_{n})\nonumber\\
&& \qquad =p_0(l^H, r^H, l^V, r^V \vert\bX)
g_T^H(h\vert\bX,l^H,r^H;\blambda^H)\nonumber
\\[-8pt]
\\[-8pt]
&& \qquad  \quad {}\times g_T^W(w\vert\bX,
h,l^V,r^V;\blambda
^W) \nn\\
&& \qquad   \quad {}\times
p_3\{\mathbf{y}\vert\bX, t_{1}-(h+w),\ldots, t_{n}-(h+w); \btheta
\}
.\nonumber
\end{eqnarray}

%s2.2.3 ###
\subsubsection{Hierarchical structure of the model}

To construct the observed data likelihood, we index each individual's
data by $i=1,\ldots, N$ and let $n_i$ be the number of observations for
the $i$th individual, ($\bY_i$, $\bX_i$, $L_i^H$, $R_i^H$, $L_i^V$,
$R_i^V$, $t_{i1},\ldots, t_{in_i})$ are observed. If we denote by
$[A\vert B; \Omega]$ the conditional distribution of random variable
$A$, given random variable $B$ and parameter $\Omega$, we can summarize
our model by a hierarchical structure from a Bayesian point of view:
%
%e2.6 ###
\begin{eqnarray}\label{fullmodel}
[\bY_i \vert\bX_i, H_i, W_i, t_{i1},\ldots, t_{in_i}; \btheta]
&\sim&
P_3 (\mathbf{y}\vert\bX_i, t_{i1}-v_i,\ldots, t_{in_i}-v_i; \btheta
), \nonumber\\
{[}W_i \vert\bX_i, H_i,L_i^V, R_i^V ; \blambda^W] &\sim& G_T^W(w
\vert
\bX
_i, h_i, l_i^V, r_i^V; \blambda^W), \nonumber\\
{[}H_i \vert\bX_i, L_i^H, R_i^H; \blambda^H] &\sim& G_T^H(h \vert\bX_i,
l_i^H, r_i^H; \blambda^H),\nonumber
\\[-8pt]
\\[-8pt]
{[}L_i^H, R_i^H, L_i^V, R_i^V \vert\bX_i] &\sim& P_0(l^H, r^H, l^V, r^V
\vert\bX_i;\bdelta), \nonumber\\
{[}\bdelta, \blambda^H,\blambda^W,\btheta]&\sim& F(\bdelta
,\blambda
^H,\blambda^W,\btheta), \nonumber\\
v_i&=&h_i+w_i, \qquad  i=1, \ldots, N, \nn
\end{eqnarray}
where $P_3(\cdot)$, $G_T^W(\cdot)$, $G_T^H(\cdot)$, $P_0(\cdot)$ and
$F(\cdot)$ are the corresponding distribution functions.
Assuming the independence of the priors for $\bdelta$ and $(\blambda
^H,\blambda^W,\btheta)$, the marginal distribution of the censoring
intervals $P_0(l^H, r^H, l^V, r^V \vert\bX_i;\bdelta)$ is not part of
the posterior inference about $(\blambda^H,\blambda^W,\btheta)$ because
of the noninformative censoring conditions. Therefore, we do not need
to model $P_0(l^H, r^H, l^V, r^V \vert\bX_i;\bdelta)$ explicitly.

%s2.2.4 ###
\subsubsection{Semiparametric Bayesian approach for event time
distributions} \label{semi}
We use a semiparametric Bayesian approach for modeling $H$ and $W$.
The CDFs $G^H$ and $G^W$ are left unspecified and not constrained to a
parametric family. Therefore, $G^H$ and $G^W$ are themselves unknown
parameters, and Dirichlet process priors [\citet{ferguson1973}]
are assigned.

A Dirichlet process prior (DPP) on a nonparametric distribution $G$
%%(we drop the superscript $H$ and $W$ of $G^H$ and $G^W$ for
%simplicity)
is a distribution on the space of all possible distributions for $G$
[\citet{ferguson1973}]. The parameters of DPP are a parametric
distribution $G_0(\cdot; \blambda)$, and a positive scalar $\alpha$.
The parametric distribution $G_0$ corresponds to the prior expectation
of the distribution function $G$. The precision parameter $\alpha$
indicates how similar we believe the base measure $G_0$ and the
nonparametric distribution $G$ are. A DPP with parameters $\alpha$ and
$G_0$ is denoted by $\mathcal{D}(\alpha G_0)$.

In the HERS analysis reported in Section~\ref{analysis2}, we include
baseline HCV status as the covariate for event time distributions.
Therefore, adding nonparametric DPP for $G^H$ and $G^W$ with base
measures $G_0^H$, $G_0^W$, and precision parameters $\alpha^H$,
$\alpha
^W$, the initial hierarchical model structure in (\ref{fullmodel}) can
be elaborated as
%
%e2.7 ###
\begin{eqnarray}\label{semimodel}
[\bY_i \vert\bX_i, H_i, W_i, t_{i1},\ldots, t_{in_i}; \btheta]
&\sim&
P_3(\mathbf{y}\vert\bX_i,t_{i1}-v_i,\ldots, t_{in_i}-v_i; \btheta)
,\nonumber\\
{[}W_i \vert\bX_i, H_i,L_i^V, R_i^V] &\sim& G_T^W(w \vert Z_i, h_i, l_i^V,
r_i^V),\\
{[}G^W(\cdot\vert Z_i); \blambda^W, \alpha^W] &\sim&\mathcal
{D}(\alpha^W
G_0^W(\cdot\vert Z_i; \blambda^W)),\nonumber\\
{[}H_i \vert\bX_i, L_i^H, R_i^H] &\sim& G_T^H(h\vert Z_i, l_i^H,
r_i^H),\nonumber
\\%[-8pt]
{[}G^H(\cdot\vert Z_i) ; \blambda^H, \alpha^H] &\sim&\mathcal
{D}(\alpha^H
G_0^H(\cdot\vert Z_i; \blambda^H))\nonumber,\\
{[}L_i^H, R_i^H, L_i^V, R_i^V \vert\bX_i] &\sim& P_0(l^H, r^H, l^V, r^V
\vert\bX_i;\bdelta), \nonumber\\
{[}\bdelta, \blambda^H,\blambda^W,\btheta]&\sim& F(\bdelta
,\blambda
^H,\blambda^W,\btheta), \nn\\
v_i&=&h_i+w_i, \qquad  i=1, \ldots, N. \nn
\end{eqnarray}

%s2.2.5 ###
\subsubsection{Model for CD4 counts}\label{modelcd4}

In this section we describe the model for CD4 counts. Recall that our
objective is to
characterize mean CD4 count profiles relative to individual viral
suppression times for HCV groups, after adjusting for other covariates.
In other words, our focus is on the parameter $\btheta$ in
$P_3(\mathbf{y}
\vert\bX_i,t_{i1}-(h_i+w_i),\ldots, t_{in_i}-(h_i+w_i); \btheta)$.
Since viral suppression time $V$ can be right-censored, those
individuals with $V$ less than or equal to the maximum follow-up time
$T$ are treated as HAART responders, while those with $V>T$ are
considered as nonresponders in the study period for comparison purpose.
It is also assumed that the mean CD4 count profiles differ by both
HAART responder groups and HCV groups; thus, different smooth functions
are used for these subpopulations. We only realign the data for the
HAART responder group by viral suppression times; for the nonresponder
group the measurement time origin is still participant
enrollment.

In addition, there are other important covariates that are possibly
associated with immunologic response to HAART besides the HCV
serostatus, for example, the overall CD4 level before HAART initiation
and baseline injection drug use information.
Specifically, let $\bX_{i}^*$ be a vector of other covariates excluding
baseline HCV status $Z_i$,
%the observed CD4 count just before HAART initiation,
and $T$ be the maximum follow-up time for the study. For $j=1, \ldots
,n_i$, we assume that the CD4 count at $t_{ij}$ for the $i$th
individual follows
%
%
%e2.8 ###
\begin{eqnarray}\label{cd4model}
Y_{ij} \vert\bX_{ij}^*, Z_i, v_i,t_{ij} =
\cases{
m_{i}(t_{ij}-v_i) + \bX_{i}^* \bbeta^* + e_{ij}, &  \quad if $v_i \le
T$,\cr
c_{i}(t_{ij}) + \bX_{i}^* \bbeta^* + e_{ij}, &  \quad if $v_i > T$,
}
\end{eqnarray}
where
\begin{eqnarray*}
m_{i}(t)&=&Z_i \cdot m_{1}(t)+(1-Z_i)\cdot m_{0}(t)+\gamma_{i}^m(t),\\
c_{i}(t)&=&Z_i \cdot c_{1}(t)+(1-Z_i)\cdot c_{0}(t)+\gamma_{i}^c(t).
\end{eqnarray*}
Here $m_{1}(t)$, $m_{0}(t)$, $c_{1}(t)$, $c_{0}(t)$ are smooth
functions describing the population CD4 count profiles that are
specific to HCV serostatus, $\gamma_{i}^m(t)$ and $\gamma_{i}^c(t)$
are individual-level smooth functions that represent random deviations
from population profiles, $\bbeta^*$ is the regression coefficient for
$\bX_{ij}^*$, and the\vspace*{-2pt} within-individual error term $e_{ij} \stackrel
{\mathrm{i.i.d.}}{\sim}N (0, \sigma^2)$. We assume that $e_i(t)$, $\gamma
_{i}^m(t)$ and $\gamma_{i}^c(t)$ are mutually independent. Detailed
specification for all smooth functions can be found in the \hyperref[appm]{Appendix}.
Overall, $m_{1}(t)$, $m_{0}(t)$, $c_{1}(t)$, $c_{0}(t)$ can be
considered as fixed effects, $\gamma_{i}^m(t)$, $\gamma_{i}^c(t)$ can
be considered as random effects, and $e_{ij}$ is the measurement error
in the linear mixed model framework. Because within-subject covariance
is not of direct interest in our analysis, no stochastic process is
further introduced into the CD4 count model except random effects and
measurement error. However, when within-subject covariance is the
target of inference, stochastic processes, for example, the integrated
Ornstein--Uhlenbeck process in \citet{Taylor1994}, can be added.

%realization of a zero mean Gaussian stochastic
%process $e_i(t)$ with covariance function $\rho(t_1, t_2) = \operatorname{cov}
%t_2$ and $\rho(t_1, t_2)=\sigma^2$ if $t_1 = t_2$. We further assume
%that $e_i(t)$, $\gamma_{i}^m(t)$ and $\gamma_{i}^c(t)$ are mutually
%independent.
%Note that $\beta^* \in\btheta$, but for simplicity in the following
%sections we suppress the notation of conditioning on $X_i$ for
%$[Y_{ij} \vert Z_i,V_i, t_{ij}]$ as in (\ref{cd4model}).

%s3 ###
\section{Prior specification and posterior inference}
\label{estimation2}

%To make inference about unknown parameters $\blambda^H$, $
%$W$ given $Z$ based on observed CD4 counts $\bY$ and censoring
%intervals,
Gibbs sampling can be used to obtain posterior samples from the full
conditional posterior distributions of $\blambda^H$, $\blambda^W$ and
$\btheta$. Compared to the model with known $H$ and $W$ in (\ref
{simple2}), the model in (\ref{semimodel}) involves an extra layer in
the Gibbs steps. That is, at each iteration, the doubly
interval-censored $W$ together with $H$ are sampled from their
conditional posterior distributions, which results in a complete data
set that is used to update the posterior distributions of the model parameters.

For the HERS analysis in Section~\ref{analysis2}, we assume that the
prior for $\btheta$ and the prior for $\blambda^H$, $\blambda^W$ are
independent. Normal distributions are used as base measures of DPP for
$G^H$ and $G^W$. Different values of the precision parameters $(\alpha
^H, \alpha^W)$ are used to evaluate the sensitivity in estimating $G^H$
and $G^W$. For the CD4 count model, standard vague priors, such as
normal-gamma conjugate family, are used.

Let
$\bH=(H_1, \ldots, H_N)\trans$, $\bW=(W_1, \ldots, W_N)\trans$,
$T_i=(t_{1i}, \ldots, t_{in_i})\trans$ and $\bT=(T_1, \ldots,
T_N)\trans
$; $\bL^H$, $\bR^H$, $\bL^V$ and $\bR^V$ are the vectors of left and
right endpoints for censoring intervals. To derive the full conditional
distribution for model (\ref{semimodel}), we use the \textit{Polya urn
characterization} of DPP [\citet{blackwell1973}] and extend the
ideas of
\citet{escobar1994} and \citet{calle2005}.
Specifically, we sample from $[ \bH, \bW, \blambda^H, \blambda^W,
\btheta\vert\bY_1,\ldots, \bY_N,\break \bX_1, \ldots, \bX_N, \bL^H,
\bR^H,
\bL^V, \bR^V, \bT]$, by the iterations as follows: first, $\bH$ and
$\bW$ are imputed by using corresponding conditional distributions;
second, the parameter $\btheta$ is updated using the complete data set
obtained from the first step and current values of the rest of
parameters; last, the parameters $\blambda^H, \blambda^W$ are updated
using distinct values of imputed $\bH$ and $\bW$. Details on priors and
full conditional posterior distributions are given in the \hyperref[appm]{Appendix}.
\looseness=1

%s4 ###
\section{Data analysis}
\label{analysis2}

\begin{figure}

\includegraphics{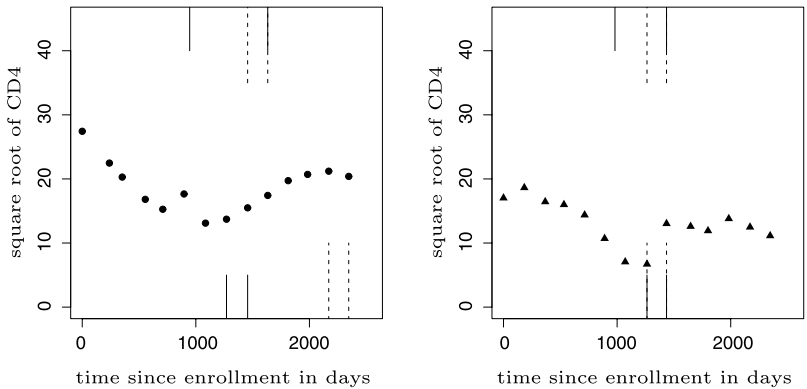}

\caption{CD4 counts (on square root scale, circles: HCV positive,
triangles: HCV negative) and censoring intervals of $H$ and $V=H+W$
under two definitions of HAART initiation time intervals for two
selected women in the HERS cohort; censoring intervals under `narrow'
definition are represented by dashed lines, censoring intervals under
`wide' definition are represented by solid lines; censoring intervals
of $H$ and $V=H+W$ are on the top and bottom of panels,
respectively.}\label{intervals}
\vspace{-10pt}
\end{figure}

In this section we apply the joint model to the HERS data introduced in
Section~\ref{hersintro}.
Two different definitions are used for censoring intervals of HAART
initiation and the results are compared. The first one is explicitly
based on reported HAART use information, and we refer to them as
`narrow' intervals for $H$. Here $R^H$ is the first visit with reported
HAART use; $L^H$ is the immediate previous visit without HAART use.
There are $159$ ($89$ HCV seropositive, $70$~HCV seronegative) patients
with right-censored viral suppression time in this case. However, we
find that some patients had viral suppression immediately before $L^H$,
which could be due to the possible reporting bias regarding HAART
initiation. As a result, we might miss the true viral suppression time
following HAART and artificially create some cases with right-censored
viral suppression time (or viral suppression that occurred long after
HAART initiation). To reduce its impact in a conservative manner,
we redefine all $374$ left endpoints of HAART initiation intervals to
be March $11$th, $1996$, which is the left endpoint of the censoring
interval for the patient who was the first reporting HAART use in the
HERS cohort. Because censoring intervals for HAART initiation are wider
under this new definition, we refer to them as `wide' intervals for $H$
and here the number of patients with right-censored viral suppression
time is reduced to $141$ ($78$ HCV seropositive, $63$ HCV
seronegative). Figure~\ref{intervals} shows the CD4 count data and
censoring intervals under two definitions of HAART initiation time
intervals for two selected women in the HERS cohort. In the left panel,
the `wide' definition for $H$ also changes the interval for viral
suppression time $V$, while in the right panel the intervals for $V$
remain the same.

%
%f4 ###

For CD4 counts, square-root transformation is used because it is
more appropriate for the assumptions of Normality and homogeneous
variance as shown by exploratory analysis.
In addition to baseline HCV serostatus, two other covariates are
included in the CD4 model: the observed CD4 count (scaled by $100$)
immediately before reported HAART initiation (pretreatment CD4 level)
and the indicator of baseline injection drug use (IDU).
For penalized splines approximating population-level smooth functions,
we use truncated quadratic bases with 20 knots, allowing sufficient
flexibility for capturing CD4 count changes at viral suppression times.
These knots are placed at viral suppression times as well as at the
sample quantiles of the realigned measurement times using midpoints of
the observed censoring intervals for viral suppression. Because data
for individual women are sparse over time and the maximum number of
data points for individual women is $15$, we use truncated quadratic
bases with one knot at the viral suppression times for estimating
individual-level smooth functions. Since the first derivatives
(velocities) of the population-level smooth functions can be computed
in analytic form when truncated quadratic bases are used, we also
examine the posterior inference for these derivatives.

%
%t2 ###
\begin{table}[b]
\tabcolsep=0pt
\caption{Percentiles (posterior mean estimates) of the time between
HAART initiation and viral suppression (in units of days) for HAART
responder group by HCV serostatus in marginal and joint models;
`narrow' stands for `narrow' intervals for $H$, `wide' stands for
`wide' intervals for $H$}
\label{times}
\begin{tabular*}{\textwidth}{@{\extracolsep{\fill}}lccc@{\hspace{3pt}}d{2.0}d{3.0}d{3.0}d{4.0}c@{}}
\hline
&&&& \multicolumn{1}{@{\hspace{3pt}}c}{$\bolds{5\%}$}
& \multicolumn{1}{c}{$\bolds{25\%}$}
& \multicolumn{1}{c}{$\bolds{50\%}$}
& \multicolumn{1}{c}{$\bolds{75\%}$}
& \multicolumn{1}{c@{}}{$\bolds{95\%}$}\\
\hline
`narrow' &$W\vert V\le T$&Marginal&HCV $+$& 15& 37& 126 &
654& 1339 \\
&&&HCV $-$& 13& 39& 118&  625& 1384 \\
&& & & & & & & \\
&&Joint&HCV $+$ & 13 &  28 &  88 & 291& \phantom{1}906 \\
&&&HCV $-$ & 13 &  31 &  82 & 356& \phantom{1}959 \\
&& & & & & & & \\
&& & & & & & & \\
`wide'&$W\vert V\le T$&Marginal&HCV $+$& 3& 145& 582 &  1129&
1497 \\
&&&HCV $-$& 1& 120& 436 &  1021& 1521 \\
&& & & & & & & \\
&&Joint&HCV $+$ & 1 &  122 &  350 & 793& 1232 \\
&&&HCV $-$&  1 &  91 &  322&  768 &  1315 \\
\hline
\end{tabular*}
\vspace{-10pt}
\end{table}

%
%t3 ###
\begin{table}[b]
\tabcolsep=0pt
\caption{Proportions (posterior mean estimates) of HAART responders and
proportions of HAART responders with time between HAART initiation and
viral suppression less than $90$ ($180$) days by HCV serostatus from
marginal and joint models in the HERS cohort; $95\%$ credible intervals
are in square brackets; `narrow' stands for `narrow' intervals for $H$,
`wide' stands for `wide' intervals for~$H$}
\label{times2}
\begin{tabular*}{\textwidth}{@{\extracolsep{\fill}}lc@{\hspace{13pt}}d{2.2}cd{2.2}@{}}
\hline
& & \multicolumn{1}{@{\hspace{13pt}}c}{$\bolds{p(V\le T)}$}
& \multicolumn{1}{c}{$\bolds{p(W\le90 \vert V\le T)}$}
& \multicolumn{1}{c@{}}{$\bolds{p(W\le180 \vert V\le T)}$}\\
\hline
`narrow' & & & & \\
 \quad Marginal&HCV $+$& 0.75&  0.42&  0.56 \\
&HCV $-$& 0.72& 0.43& 0.56\\
&Difference& -0.03& 0.02& -0.01\\
& & \multicolumn{1}{c}{[$-$0.14,\ 0.08]}& \multicolumn{1}{c}{[$-$0.24,\ 0.25]} & \multicolumn{1}{c@{}}{[$-$0.12,\ 0.11]} \\
& & & & \\
 \quad Joint&HCV $+$& 0.63&  0.48&  0.66 \\
&HCV $-$& 0.64& 0.52& 0.62\\
&Difference& 0.01& 0.05& -0.04\\
& & \multicolumn{1}{c}{[$-$0.05,\ 0.06]}& \multicolumn{1}{c}{[$-$0.24,\ 0.31]} & \multicolumn{1}{c@{}}{[$-$0.14,\ 0.07]} \\
& & & & \\
`wide' & & & & \\
 \quad Marginal&HCV $+$& 0.85& 0.13&  0.29 \\
&HCV $-$& 0.78& 0.22& 0.33\\
&Difference& -0.07& 0.08& 0.04\\
& & \multicolumn{1}{c}{[$-$0.22,\ 0.07]}& \multicolumn{1}{c}{[$-$0.03,\ 0.19]} & \multicolumn{1}{c@{}}{[$-$0.06,\ 0.14]} \\
& & & & \\
 \quad Joint&HCV $+$& 0.68&  0.17&  0.36 \\
&HCV $-$& 0.68&  0.24&  0.38\\
&Difference& 0.01& 0.07& 0.01\\
& & \multicolumn{1}{c}{[$-$0.05,\ 0.07]}& \multicolumn{1}{c}{[$-$0.06,\ 0.20]} & \multicolumn{1}{c@{}}{[$-$0.10,\ 0.12]} \\
\hline
\end{tabular*}
\vspace{-10pt}
\end{table}

The prior specifications are as described in Section~\ref{estimation2}
and the \hyperref[appm]{Appendix}. For assessing sensitivity in estimating $G^H$ and
$G^W$, precision parameters $(\alpha^H,\alpha^W)$ of the Dirichlet
process are taken to be equal to $(1,1)$ and $(10,10)$, which indicate
different levels of faith in the prior normal base measures for $H$ and
$W$. We run two MCMC chains with $7000$ iterations, the first $2000$ of
which are discarded. Convergence is established graphically using
history plots; pooled $10\mbox{,}000$ posterior samples are then used for
inference. The results at both values of $\alpha^H$, $\alpha^W$ are
similar; here we present those with $(\alpha^H,\alpha^W)=(10,10)$. MCMC
is implemented in MATLAB programs [\citet{matlab}].

For the purpose of modeling doubly interval-censored event time $W$
only, marginal models can be used by excluding the part for CD4 counts
from (\ref{semimodel}). We will compare the results from our joint
model with those from marginal models, and investigate the possible
impact of joint modeling.

%s4.1 ###
\subsection{Results for virologic response to HAART}

Table~\ref{times} presents the posterior mean estimates of the
percentiles of the time between HAART initiation and viral suppression
for the HAART responder group.
The results based on `wide' intervals for $H$ suggest that the HCV
negative group might have shorter time to achieve viral suppression
than the HCV positive group, but this is not the case with `narrow'
intervals for $H$, where the HCV negative group has more right skewed
distribution. Further, the joint model tends to give smaller estimates
than the marginal model. For example, in Table~\ref{times} both
location estimates and variability estimates from the joint model based
on `wide' intervals for $H$ are smaller than those from the marginal
model, which suggests that modeling CD4 counts affects the estimation
for doubly interval-censored $W$ when the information from censoring
intervals is limited.

\begin{figure}

\includegraphics{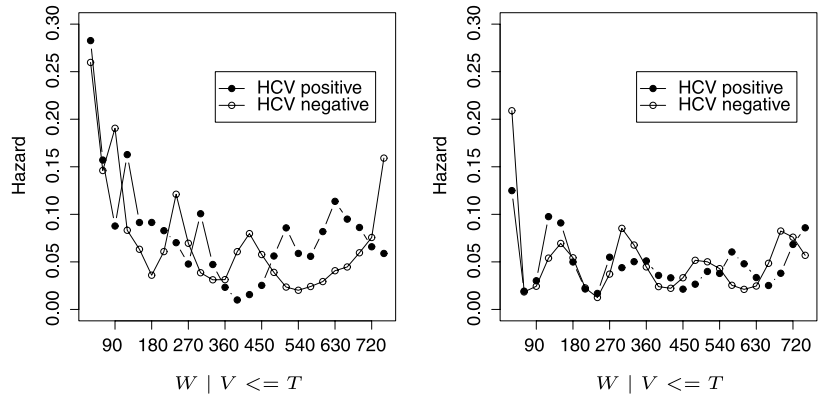}

\caption{Hazard function of viral suppression after HAART initiation by
HCV serostatus in the HERS cohort over grid points of $30$ days from
the joint model; left panel: `narrow' intervals for $H$; right panel:
`wide' intervals for $H$.}\label{HT}
\vspace{-10pt}
\end{figure}

Table~\ref{times2} gives the estimated proportions of HAART responders
with time between HAART initiation and viral suppression less than or
equal to $90$/$180$ days. In both cases of `wide' and 'narrow'
intervals for $H$, the $95\%$ credible intervals for differences
between proportions by HCV groups cover zero.
Thus, in the HERS cohort, there is not sufficient evidence that
baseline HCV serostatus is associated with virologic response to HAART.
This is also demonstrated in Figure~\ref{HT}, where the hazard
functions of viral suppression are plotted over grid points of $30$
days. Here the hazard is defined as $p(W< t_2 \vert W \ge t_1, V\le T)$,
where $t_1$, $t_2$ are grid points. With both `narrow' and `wide'
intervals for $H$, the hazard functions of viral suppression are
generally similar across the HCV groups. Note that estimated
proportions of HAART responders $p(V\le T)$ are also similar for the
HCV groups in all cases.

%
%f5 ###

From Table~\ref{times}, median estimates for the time between HAART
initiation and viral suppression are approximately one year with `wide'
intervals for $H$ and $3$--$4$~months with `narrow' intervals for $H$
in the joint model. Compared to the clinically expected value, the
estimates with `wide' intervals for $H$ might be overestimated due to
the following reasons. First, data were collected approximately every
six months in the HERS, thus the immediate virologic response to HAART
were not available. Second, HAART information was self-reported and we
set up the left endpoints of HAART initiation time to be March 11th,
1996 for reducing reporting bias. Consequently, censoring intervals for
observed HAART initiation times are wide. Third, $38\%$ of the
participants had right-censored viral suppression times, which might be
related to the adherence of HAART treatment and individual
heterogeneity in virologic response. However, these situations do not
differ by HCV serostatus, thus the corresponding comparison can still
be useful.

%s4.2 ###
\subsection{Results for immunologic response to HAART}

The results for CD4 counts are similar under both definitions of
censoring intervals for HAART initiation and we present those based on
`wide' intervals for $H$.

%s4.2.1 ###
\subsubsection{Population estimates}

We compute posterior mean estimates for all targets of inference. The
coefficient estimate for pretreatment CD4 level is $2.35$ ($95\%$
credible interval $[2.22, 2.49]$), which clearly indicates the positive
association between pretreatment CD4 level and the current CD4 count,
given baseline HCV and IDU statuses. The coefficient estimate for
baseline IDU is $-0.06$ ($95\%$ credible interval $[-0.80, 0.64]$),
suggesting that baseline IDU status was not associated with current CD4
counts, given baseline HCV and pretreatment CD4 level.

%f6 ###
\begin{figure}[b]

\includegraphics{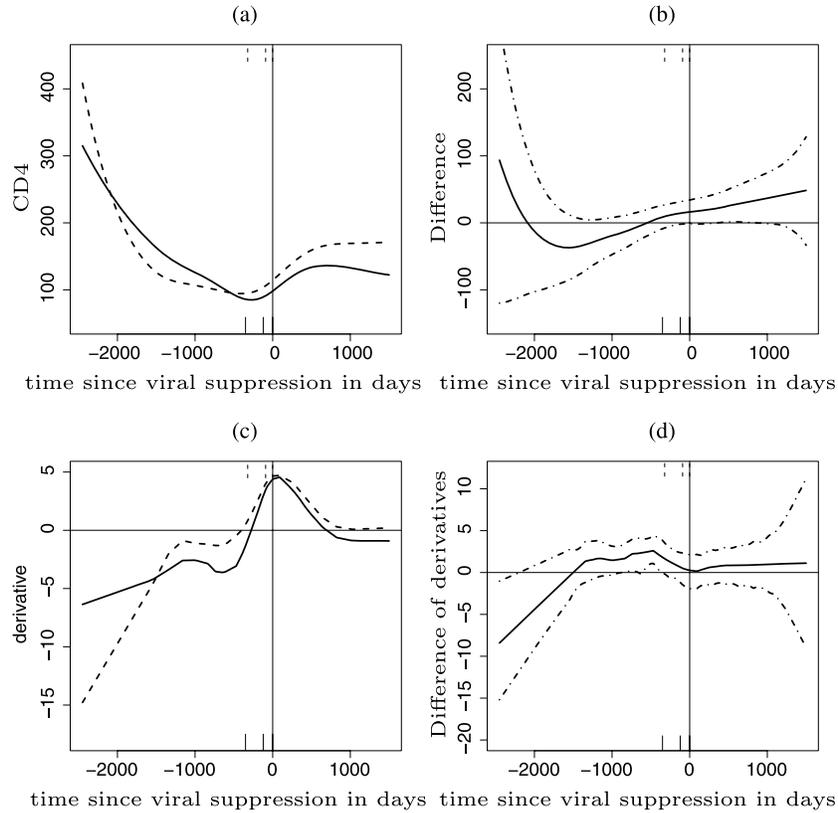}

\caption{\textup{(a)} Estimated CD4 count profiles by HCV groups for HAART
responders (transformed to original CD4 count scale) in the joint
model, after accounting for pretreatment CD4 level and baseline
injection drug use: solid line, HCV-positive group; dotted line, HCV-negative group.
\textup{(b)}~Difference between CD4 count profiles (in original
CD4 count scale) in the joint model: solid line, posterior mean
estimates; dotted lines, $95\%$ pointwise credible bands. \textup{(c)}
Derivatives for CD4 count profiles by HCV groups for HAART responders
(in square root CD4 count scale) in the joint model, after accounting
for pretreatment CD4 level and baseline injection drug use. \textup{(d)}
Difference between derivatives for CD4 count profiles (in square root
CD4 count scale) in the joint model. The ticks at the top and the
bottom of the panels are the HAART initiation times corresponding to
the $5\%$, $50\%$ and $95\%$ quantiles of the time between HAART
initiation and viral suppression in Table~\protect\ref{times}: solid
line, HCV-positive group; dotted line, HCV-negative group.}\label{herspop}
\vspace{-10pt}
\end{figure}

For HAART responders, mean CD4 count profiles (after accounting for
pretreatment CD4 level and baseline IDU) are plotted in the panel (a)
of Figure~\ref{herspop}. We transform the estimates back to the
original CD4 count scale for illustration purposes. The estimated CD4
count profiles of both HCV groups were decreasing at $3$--$6$ years
before viral suppression. CD4 counts started to increase before HIV
virus was completely suppressed (time point $0$). This is consistent
with findings from other studies, that is, CD4 cells may increase after
HAART for patients who do not fully suppress the virus, because the
level of viral load is decreasing [\citet{jacobson2004}]. However,
Figure~\ref{herspop}(a), also suggests that the decreasing
trend for HCV-negative patients ends earlier than HCV-positive patients
when HAART started to be initiated. In addition, the average CD4 level
after viral suppression achieved by HCV-negative patients is higher
than HCV-positive patients. For example, at viral suppression time the
difference of average CD4 count for HCV groups is approximately $16$
($95\%$ credible interval $[-3, 35]$), controlling for pretreatment CD4
level and baseline IDU. We also plot the difference curve between mean
CD4 count profiles of HCV groups [Figure~\ref{herspop}(b), in
original CD4 count scale]. The pointwise $95\%$ credible bands are
approximately above zero after CD4 counts started to increase. Note
that the difference between point estimates of the mean CD4 counts at
the left boundary for the time since viral suppression axis might be
due to the small sample size and large estimation variability, which is
suggested by the width of $95\%$ pointwise credible bands.

%
%f7 ###
\begin{figure}[b]

\includegraphics{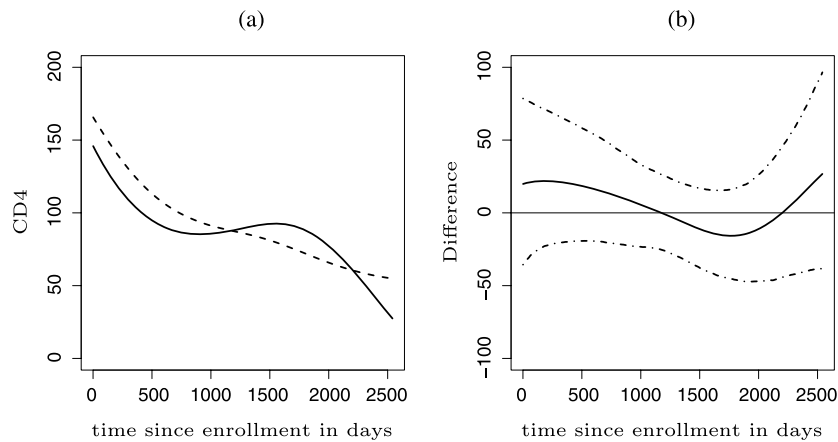}

\caption{\textup{(a)} Estimated CD4 count profiles by HCV groups for HAART
nonresponders (transformed to original CD4 count scale) in the joint
model, after accounting for pretreatment CD4 level and baseline
injection drug use: solid line, HCV positive group; dotted line, HCV
negative group. \textup{(b)}~Difference between CD4 count profiles (in original
CD4 count scale) in the joint model: solid line, posterior mean
estimates; dotted lines, $95\%$ pointwise credible bands.}\label{popnore}
\vspace{-10pt}
\end{figure}

To evaluate immune reconstitution after HAART, the rate of CD4 count
change is a useful measure. Panel (c) of Figure~\ref{herspop} presents
the derivative (velocity) curves for mean CD4 count profiles of HAART
responders. For both HCV groups, the velocities of the average CD4
count change reach the maximum
approximately at viral suppression times, which is sensible because the
major driving force of immune reconstitution is viral suppression
[\citet
{jacobson2004}]. Overall, the HCV-negative group has slightly larger
point estimates of mean CD4 count change rate leading up to and
following viral suppression.
Panel (d) of Figure~\ref{herspop} gives the difference and the
corresponding $95\%$ credible bands between derivative curves of HCV
groups. After controlling for pretreatment CD4 level and baseline IDU,
the rates of mean CD4 count change do not appear to be different by HCV
serostatus in the HERS cohort.

The left panel of Figure~\ref{popnore} presents the mean CD4 count
profiles for HAART nonresponders (in original CD4 count scale) along
the time since enrollment. Both HCV groups had the same decreasing
patterns, and the difference curve and its $95\%$ credible band (right
panel of Figure~\ref{popnore}) indicate that there is not difference in
mean CD4 count levels for HCV groups in this nonresponder population,
after adjusting for pretreatment CD4 level and baseline IDU.

%s4.2.2 ###
\subsubsection{Individual estimates}

\begin{figure}

\includegraphics{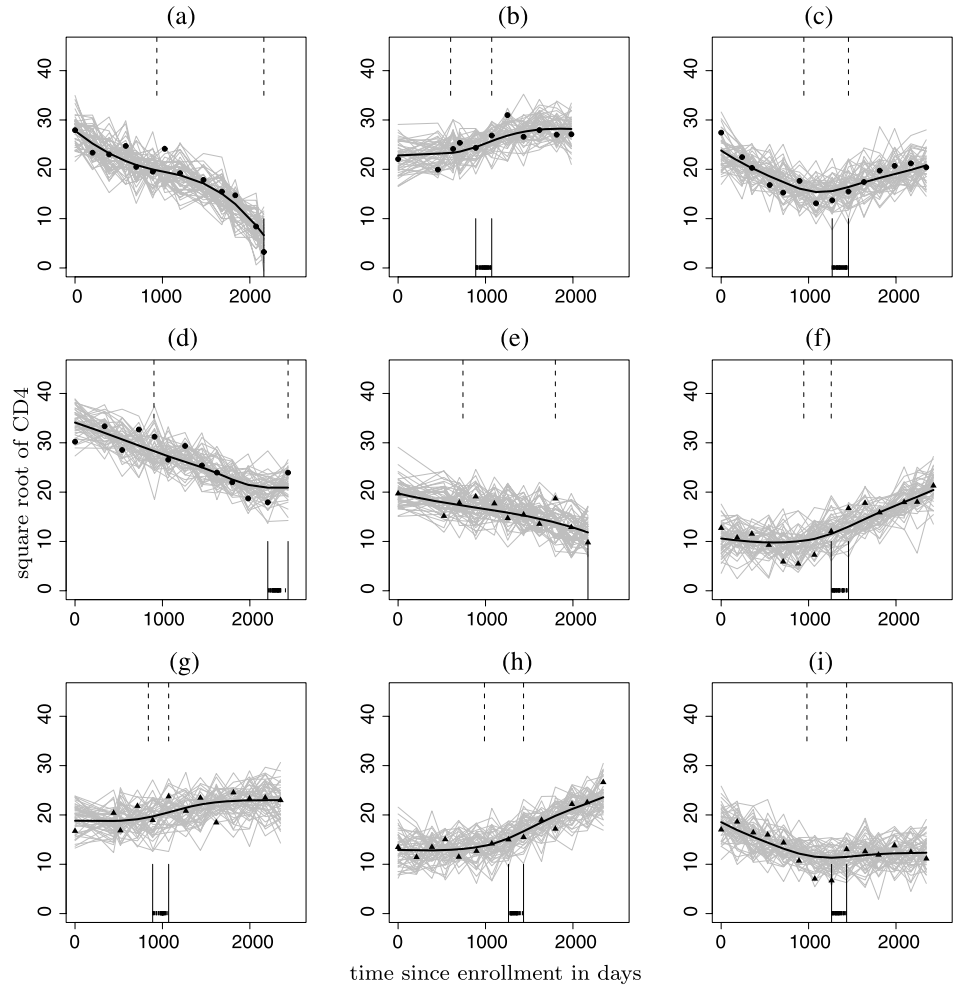}

\caption{CD4 count data (on square root scale) and $50$ posterior
predictive sample curves in the joint model from $9$ selected women in
the HERS cohort: vertical dotted lines are censoring intervals for
HAART initiation (under `wide' definition), vertical solid lines are
censoring intervals for viral suppression; except for panels \textup{(a)} and
\textup{(e)} with $v_i>T$, ticks at the bottom of each panel are imputed viral
suppression times ($v_i\le T$); circles represent data from the HCV-positive group and
triangles represent data from the HCV-negative
group; solid lines are averages of $50$ sampled mean curves.}\label{indi}
\vspace{-10pt}
\end{figure}

The parameter estimates for
individuals may not exactly follow the patterns of the population if
the between-subject variation is large.
Data, $50$ sample curves from posterior predictive distributions and
averages of $50$ sampled mean curves for nine selected HERS women in
Section~\ref{introduction2}, are plotted in Figure~\ref{indi}.
Compared with Figures~\ref{herspop} and \ref{popnore}, we can see that
not only the magnitude but also the patterns are different between the
population and individual estimated profiles. However, the model fits
well to this representative sample of individuals.

%
%f8 ###

%s5 ###
\section{Conclusion and discussion}
\label{conclusion2}

We proposed a joint model for doubly interval-censored event time and
longitudinal data in HIV natural history studies in order to
investigate the post-HAART HIV dynamics and the associated factors.
Using data from the HERS cohort, we found that HCV-negative and HCV-positive patients had similar virologic response, which is measured by
the time from HAART initiation to viral suppression. Further, our
results show that for patients with virologic response to HAART, being
HCV seronegative is associated with higher average CD4 count level
after viral suppression, given the same pretreatment CD4 level and
baseline IDU status. The HCV-negative group showed slightly higher
immune reconstitution level (measured by the rate of mean CD4 count
change) leading up to and following viral suppression, however, the
evidence from the HERS cohort is not sufficient to support the conclusion.

Data from natural history studies have been used to evaluate the effect
of HCV coinfection on post-HAART HIV dynamics [\citet
{greub2000}; \citet{sulk2002}; \citet{miller2005}].
However, virologic response and
immunologic response were investigated separately and simple summary
statistics were used for inference, for example, average CD4 count
increases after HAART initiation by visits, hazard ratio of increasing
CD4 count by at least $50$ cells/\textmu l in a year, etc. In contrast,
our method considers the characteristics of longitudinal cohort data as
well as the biological background of the post-HAART HIV dynamics (such
as the sequential relationship between virologic and immunologic
response); our joint modeling approach
utilizes all available information from natural history studies and the
results can be informative in generating hypotheses for AIDS clinical trials.

In the HERS analysis, we considered the women with $V>T$ as HAART
nonresponders and examined their population mean CD4 count profiles.
However, because the data are from a natural history study and the
observed HAART initiation times vary across individuals, the observed
data for viral suppression time actually depend on the timing of HAART
initiation. Therefore, the HERS women with $V>T$ might not be a
homogenous group in terms of response to HAART. The definition of
`responder,' however, does not differ by HCV status. Thus, for
comparison purposes, it would still be useful to examine the population
mean CD4 count profiles for both women with $V>T$ and women with $V \le T$.

Due to the sparse data, information on event times for evaluating
virologic response is limited in the HERS cohort. In order to reduce
possible reporting bias regarding HAART initiation, we use two
definitions of censoring intervals for HAART initiation and investigate
the impact on the analysis. The conclusions for HCV serostatus and
post-HAART HIV dynamics do not differ by the definitions. However, the
actual estimates for time between HAART initiation and viral
suppression might be larger compared to the clinical expected values
due to the study design, conservative definition of censoring
intervals, participant noncompliance, drug resistance and other
individual heterogeneity in virologic response to HAART.
As we are being conservative by moving left endpoints of HAART
initiation time to the earliest possible date, another option could be
a hybrid approach by changing censoring intervals only for those with
suspicious viral suppression immediately before self-reported HAART
initiation date. Alternatively, we could specify a uniform prior for
the left boundary of HAART initiation time between the left boundaries
defined in `narrow' and `wide' intervals to reflect uncertainty about
true HAART initiation time.

Besides HCV coinfection, other potential determinants
or modifiers of post-HAART HIV dynamics include
characteristics of the HAART regimen, prior antiviral treatment
history, stage of disease at the time of HAART initiation (viral load
level), an
intact immune system and other host characteristics, such as age, race,
gender and genotype [\citet{jacobson2004}]. For adjusting these possible
factors, covariates can be added into the CD4 count model (\ref
{cd4model}) similarly as for the case of pretreatment CD4 level and
baseline IDU status. For doubly interval-censored data, one limitation
of our Bayesian semiparametric approach is that sample sizes could be
small for reliable estimation when the unique values of the covariates
are large. For example, there were only 4 HERS women who were IDU and
HCV negative at baseline. Therefore, we could not assign different DPP
to all combinations of the covariate values when baseline IDU is
included as a covariate. In this scenario, a parametric approach can be
developed to adjust for additional covariates.

We believe that the proposed joint modeling approach is methodologically
valuable. The proposed regression spline method is simple to implement, and
naturally incorporates the typical features of longitudinal data such
as between-individual and within-individual
variations. The proposed model can be extended to characterize multiple
processes in disease progression after treatment intervention, for
example, the neurocognitive response to HAART treatment after immune
reconstitution is another process of interest apart from the virologic
and immunologic response [\citet{bell2004}].

\begin{appendix}

\section*{\texorpdfstring{Appendix: Full conditional distributions for Gibbs
steps~in~Section~\lowercase{\protect\ref{estimation2}}}{Appendix: Full conditional distributions for Gibbs
steps~in~Section~3}}
\label{appm}

%s5.1 ###
\subsection{Data augmentation for event times}

A value for each censored observation, $H_i$, is sampled from the
conditional distribution of $H_i$ given all other parameters. Under a
DPP this conditional distribution maintains the same Polya urn
structure assumed \textit{a priori} for $H_1, \ldots, H_N$. It can be
shown that the full conditional distribution of $H_i$ has the following form:
%
%
%e5.1 ###
\begin{eqnarray}\label{postU}
&&[H_i \vert\bY_i, T_i,\bX_i, \bW, \{H_j, j\ne i\}, \bL^H, \bR^H,
\bL
^V, \bR^V, \blambda^H , \blambda^W,\btheta] \nonumber
\\[-8pt]
\\[-8pt]
&& \qquad  \sim r_0
\cdot g_{0T}^H(h_i\vert Z_i,v_i, l_i^H, r_i^H, \blambda^H) + \sum_{j
\ne
i} r_j \cdot I(h_j=h_i), \nonumber
\end{eqnarray}
where $g_{0T}^H$ is the truncated posterior distribution in the
censoring interval $(L_i^H,  \operatorname{min}(R_i^H,V_i)]$. Note that
$\bY_i$ does not get involved in (\ref{postU}) because conditioning on
$V_i$, $\bY_i$ and $H_i$ are independent. Since $V_i$ only provides
information on the range of $H_i$, $g_{0T}^H$ is simply the truncated
$g_{0}^H$, base measure of $H_i$ given $Z_i$. Furthermore,
\begin{eqnarray*}
r_0 &\propto&\alpha^H \int_{l_i^H}^{ \operatorname{min}(r_i^H,v_i)}
g_0^H(h_i \vert Z_i; \blambda^H)\, dh_i, \\
r_j &\propto& I \bigl(l_i^H < h_j \le\operatorname{min}(r_i^H, v_i),
Z_j=Z_i\bigr),
\end{eqnarray*}
and $r_0+\sum_{j\ne i} r_j=1$. Thus, a new value of $H_i$ is equal
either to $h_j$ with probability $r_j$, or to a sampled value from the
distribution $g_{0T}^H$ with probability $r_0$. Also, we assume that
depending on the value of $Z_i$, the base measure $g_0^H$ are normal
distributions with distinct parameters $(\mu_1^H, \tau_1^H)$ or $(\mu
_0^H, \tau_0^H)$.

For $W_i=V_i-H_i$, the full conditional distribution follows:
\begin{eqnarray}\label{postV}
&&[W_i \vert\bY_i, T_i,\bX_i, \bH, \{W_j, j\ne i\}, \bL^H, \bR^H,
\bL
^V, \bR^V, \blambda^H , \blambda^W,\btheta] \nn\\
&& \qquad \sim q_0 \cdot
g_{0T}^W(w_i\vert\mathbf{y}_i, T_i, \bX_i, h_i, l_i^V, r_i^V,
\blambda^W) +
\sum_{j \ne i} q_j \cdot I(w_j=w_i), \nonumber
\end{eqnarray}
where
\begin{eqnarray}
&&g_{0T}^W(w_i\vert\mathbf{y}_i, T_i, \bX_i, h_i, l_i^V, r_i^V,
\blambda
^W)\nn
\\
&& \qquad \propto p_3\bigl(\mathbf{y}_i \vert\bX_i, T_i-(h_i+w_i); \btheta
\bigr)g_0^W(w_i\vert
Z_i; \blambda^W)\nn\\
&& \qquad  \quad {}\times I\bigl(\operatorname{max}(0, l_i^V-h_i)<w_i\le r_i^V-h_i\bigr) \nonumber
\end{eqnarray}
is the truncated posterior distribution of $W_i$ in $(\operatorname{max}(0, L_i^V-H_i),R_i^V-H_i]$. Furthermore,
\begin{eqnarray*}
q_0&\propto&\alpha^W \int_{\operatorname{max}(0, l_i^V-h_i)}^{r_i^V-h_i}
p_3\bigl(\mathbf{y}_i \vert\bX_i,T_i-(h_i+w_i); \btheta\bigr)g_0^W(w_i\vert Z_i;
\blambda
^W)\, dw_i,
\\
q_j&\propto& p_3\bigl(\mathbf{y}_i \vert\bX_i, T_i-(h_i+w_j); \btheta
\bigr)I\bigl(\operatorname{max}(0, l_i^V-h_i)<w_j\le r_i^V-h_i, Z_j=Z_i\bigr),
\end{eqnarray*}
and $q_0+\sum_{j\ne i} q_j=1$. Thus, a new value of $W_i$ is equal
either to $w_j$ with probability $q_j$, or to a sampled value from the
distribution $g_{0T}^W$ with probability $q_0$, where $g_{0T}^W$ is the
full conditional distribution of $W$ that would be obtained if the
completely parametric hierarchical model (\ref{fullmodel}) is used and
$g_0^W$ is the prior distribution (base measure) for $W$ given $Z$.
We again assume that $g_0^W$ are normal distributions with distinct
parameters $(\mu_1^W, \tau_1^W)$, $(\mu_0^W, \tau_0^W)$. Because
$p_3(\mathbf{y}_i \vert\bX_i,T_i-(h_i+w_i); \btheta)$ is based on
the model
in (\ref{cd4model}), there is no closed form for $g_{0T}^W$ and the
Metropolis step [\citet{gelman2003}] is used for sampling. The integral
in $q_0$ is approximated by the Gauss--Legendre quadrature with $20$ nodes.

%s5.2 ###
\subsection{Update parameters in the CD4 count model}

We use Bayesian penalized splines [\citet{ruppert2003}] with a truncated
polynomial basis for approximating CD4 count profiles at both
population level and individual level.

Following \citet{ruppert2003}, $m_1(t)$, $m_0(t)$, $c_1(t)$, $c_0(t)$,
$\gamma_i^m(t)$ and $\gamma_i^c(t)$ ($i=1, \ldots, N$) in (\ref
{cd4model}) can be approximated by
\begin{eqnarray}
\label{ft}
m_1(t)&=&\bB(t)\trans\bbeta_1,  \qquad  m_0(t)=\bB(t)\trans
\bbeta_2, \nonumber\\
c_1(t)&=&\bA(t)\trans\balpha_1,  \qquad  c_0(t)=\bA(t)\trans
\balpha_2, \nonumber\\
\gamma_i^m(t)&=&\bphi(t)\trans\mathbf{b}_i,  \qquad  \gamma_i^c(t)=\bpsi
(t)\trans\mathbf{a}_i, \nonumber
\end{eqnarray}
where $\bB(t)=(1,t, \ldots, t^p,(t-\nu_1)_+^p, \ldots, (t-\nu
_{K_B})_+^p)\trans$, $\bA(t)=(1,t, \ldots, t^p,(t-\xi_1)_+^p,
\ldots,
(t-\xi_{K_A})_+^p)\trans$, $\bphi(t)=(1,t, \ldots, t^p,(t-\eta_1)_+^p,
\ldots, (t-\eta_{K_{\phi}})_+^p)\trans$ and $\bpsi(t)=(1,t, \ldots,
t^p,(t-\zeta_1)_+^p, \ldots, (t-\zeta_{K_{\psi}})_+^p)\trans$ are
truncated polynomial bases; $p \ge1$ is an integer and $(d)_+^p=d^p
\cdot\mathrm{I}(d \ge0)$. $(\nu_1, \ldots, \nu_{K_B})$, $(\xi_1,
\ldots
, \xi_{K_A})$, $(\eta_1, \ldots, \eta_{K_\phi})$ and $(\zeta_1,
\ldots,
\zeta_{K_{\psi}})$ are the corresponding knots; ($K_B$, $K_A$,
$K_\phi
$, $K_\psi$) are the number of knots.

Let
\begin{eqnarray}
 \bbeta_1&=&(\beta_{1,0}, \ldots, \beta_{1,p+K_B})\trans,  \qquad
 \bbeta_2=(\beta_{2,0}, \ldots, \beta_{2,p+K_B})\trans,
\nonumber\\
\balpha_1&=&(\alpha_{1,0}, \ldots, \alpha
_{1,p+K_A})\trans
, \qquad  \balpha_2=(\alpha_{2,0}, \ldots, \alpha
_{2,p+K_A})\trans,\nonumber\\
\mathbf{b}_i&=&(b_{i,0}, \ldots,
b_{i,p+{K_\phi
}})\trans, \qquad
\mathbf{a}_i=(a_{i,0}, \ldots, a_{i,p+{K_\psi}})\trans,\nonumber
\end{eqnarray}
and $x_{ij}=t_{ij}-v_i$, then the proposed model in (\ref{cd4model})
can be rewritten as
\begin{eqnarray*}
&&Y_{ij} \vert\bX_i^*, Z_i, v_i, t_{ij}\\
&& \qquad =
\cases{
\bB(x_{ij})\trans\bbeta_1+\bphi(x_{ij})\trans\mathbf{b}_i+ \bX_{i}^*
\bbeta
^* + e_{ij},  &\quad if  $v_i \le T,Z_i=1$,\cr
\bB(x_{ij})\trans\bbeta_2 +\bphi(x_{ij})\trans\mathbf{b}_i+ \bX_{i}^*
\bbeta
^* + e_{ij} ,  &\quad {if} $v_i \le T,Z_i=0$,\cr
\bA(t_{ij})\trans\balpha_1 +\bpsi(t_{ij})\trans\mathbf{a}_i + \bX_{i}^*
\bbeta^* +e_{ij},&\quad {if} $v_i > T,Z_i=1$,\cr
\bA(t_{ij})\trans\balpha_2 +\bpsi(t_{ij})\trans\mathbf{a}_i+ \bX_{i}^*
\bbeta^* +e_{ij}, &\quad {if} $v_i > T,Z_i=0$.
}
\nn
\end{eqnarray*}
We use the standard prior distributions for all parameters in the CD4
count model as follows: $\bbeta^* \propto1$, for $s=0,\ldots, p$,
$p(\beta_{1,s}) \propto1$, $p(\beta_{2,s}) \propto1$, $p(\alpha
_{1,s}) \propto1$, $p(\alpha_{2,s}) \propto1$, $b_{i,s}\sim N(0,
\sigma_{b_s}^2)$, $a_{i,s}\sim N(0, \sigma_{a_s}^2)$, $\sigma_{b_s}^2
\sim \operatorname{Gamma}(10^{-3}, 10^{-3})$ and $\sigma_{a_s}^2 \sim \operatorname{Gamma}(10^{-3},
10^{-3})$; for $k=1, \ldots, K_B$, $\beta_{1,p+k}\sim N(0, \sigma
_{\beta
_1}^2)$ and $\beta_{2,p+k}\sim N(0, \sigma_{\beta_2}^2)$; for $k=1,
\ldots, K_A$, $\alpha_{1,p+k}\sim N(0, \sigma_{\alpha_1}^2)$ and
$\alpha
_{2,p+k}\sim N(0, \sigma_{\alpha_2}^2)$; for $k=1,\ldots, K_\phi$,
$b_{i,p+k}\sim N(0, \sigma_{b}^2)$; for $k=1,\ldots, K_\psi$,
$a_{i,p+k}\sim N(0, \sigma_{a}^2)$; $\sigma_{\beta_1}^2$, $\sigma
_{\beta
_2}^2$, $\sigma_{\alpha_1}^2$, $ \sigma_{\alpha_2}^2$, $\sigma_{b}^2$,
$\sigma_{a}^2$ all follow $\operatorname{Gamma}(10^{-3}, 10^{-3})$ distribution.
Note that $\sigma_{\beta_1}^2$, $\sigma_{\beta_2}^2$, $\sigma
_{\alpha
_1}^2$, $\sigma_{\alpha_2}^2$ are smoothing parameters for the
population penalized splines; $\sigma_{b}^2$ and $\sigma_{a}^2$ are
smoothing parameters for individual penalized splines; $\sigma
_{b_s}^2$, $\sigma_{a_s}^2$ ($s=0,\ldots, p$) are variance component
parameters for random effects.
Further, we assume $e_{ij}\sim N(0,\sigma^2)$ for all observations and
$\sigma^2 \sim \operatorname{Gamma}(10^{-3}, 10^{-3})$.

Thus, the parameter vector $\btheta$ includes ($\bbeta^*, \bbeta_1,
\bbeta_2, \balpha_1, \balpha_2, \mathbf{b}_i, \mathbf{a}_i$) and
($\sigma
_{\beta
_1}^2$, $\sigma_{\beta_2}^2$, $\sigma_{\alpha_1}^2$, $\sigma
_{\alpha
_2}^2$, $\sigma_{b}^2$, $\sigma_{a}^2$, $\sigma_{b_s}^2$, $\sigma
_{a_s}^2$, $\sigma^2$). Since all conditional posterior distributions
for $\btheta$ are in closed form, the Gibbs steps are straightforward.

%s5.3 ###
\subsection{Update parameters for DPP base measures $G_0^H$ and $G_0^W$}
The parameters $\blambda^H$ and $\blambda^W$ are updated from their
full conditional distributions:
\begin{eqnarray}
&&[\blambda^H \vert\bY_1, \ldots, \bY_N, \bX_1, \ldots, \bX_N, \bT
, \bH,
\bW, \bL^H, \bR^H, \bL^V, \bR^V, \btheta, \blambda^W] \nn\\
&& \qquad \sim\prod_{i \in\bI^H} g_0^H(h_i\vert Z_i,v_i, l_i^H, r_i^H;
\blambda
^H)f(\blambda^H), \nonumber
\\
&&\blambda^W \vert\bY_1, \ldots, \bY_N, \bX_1, \ldots, \bX_N, \bT
, \bH,
\bW, \bL^H, \bR^H, \bL^V, \bR^V, \btheta, \blambda^H] \nn\\
&& \qquad \sim
\prod
_{i \in\bI^W} g_0^W(w_i\vert Z_i, h_i, l_i^V, r_i^V; \blambda
^W)f(\blambda^W), \nonumber
\end{eqnarray}
where $\bI^H$ and $\bI^W$ are the subsets of indexes corresponding to
the distinct $H_i$ and $W_i$ because the distinct $H_i$ and $W_i$ are
random samples from $G_0^H$ and $G_0^W$, respectively [\citet
{blackwell1973}]. In our case, $\blambda^H=(\mu_1^H,\mu_0^H, \tau_1^H,
\tau_0^H)$ and $\blambda^W=(\mu_1^W, \mu_0^W,\tau_1^W, \tau_0^W)$ for
the normal base measures; we assume $f(\mu_1^H, \mu_0^H,\tau_1^H,
\tau
_0^H)\propto(\tau_1^H \tau_0^H)^{-1}$ and $f(\mu_1^W,\mu_0^W, \tau
_1^W, \tau_0^W)\propto(\tau_1^W \tau_0^W)^{-1}$. The conditional
posterior distributions of $\blambda^H$ and $\blambda^W$ are both in
closed forms.
\end{appendix}

\section*{Acknowledgments}
We are grateful to Jeffrey Blume, Mike Daniels, Constantine Gatsonis,
Patrick Heagerty, Tony Lancaster and the referees for helpful comments.
   Data
for HERS were collected under CDC cooperative agreements U64-CCU106795,
U64-CCU206798, U64-CCU306802 and U64-CCU506831.

%suskaldyti doi

\printaddresses

\end{document}